\documentclass[a4paper,11pt]{article}
\usepackage{jheppub,bm}
\usepackage[T1]{fontenc} 
\usepackage{slashed}
\usepackage{soul}
\usepackage{amsmath}
\newcommand{\bef}{\begin{figure}[hbt]\centering}
\newcommand{\eef}{\end{figure}}
\allowdisplaybreaks

\usepackage{epsfig}
\usepackage{amsmath}
\input{epsf}
\usepackage{psfrag}

\usepackage{graphicx}
\def\bea#1\eea{\begin{align}#1\end{align}}

\def\scnu{Guangdong Provincial Key Laboratory of Nuclear Science, Institute of Quantum Matter, South China Normal University, Guangzhou 510006, China}
\def\sdu{Key Laboratory of Particle Physics and Particle Irradiation (MOE), Institute of Frontier and Interdisciplinary Science, Shandong University (QingDao), Shandong 266237, China}

\title{The $\cos 2\phi$ azimuthal asymmetry in   $\rho^0$  meson production in ultraperipheral heavy ion collisions}

\author{Hongxi Xing$^{a}$, }
\author{Cheng Zhang$^{b}$, }
\author{Jian Zhou$^{b}$}
\author{and Ya-Jin Zhou$^{b}$}

\affiliation[a]{\scnu}
\affiliation[b]{\sdu}

\abstract{
We present a detailed study of vector meson photoproduction in
ultraperipheral heavy ion collisions (UPCs). Using the dipole model,
we develop a framework for the joint impact parameter and transverse
momentum dependent cross sections. We compute the unpolarized
 cross section and $\cos 2\phi$ azimuthal angular
correlation for $\rho^0$ photoproduction with $\phi$ defined as the
angle between the $\rho^0$'s transverse transverse momentum
 and its decay product pion meson's transverse momentum. Our result on unpolarized coherent differential cross
section gives excellent description to the STAR experimental data. A
first comparison between theoretical calculation and experimental
 measurement on the $\cos 2\phi$ azimuthal asymmetry, which results from the linearly polarized
 photons, is performed and reasonable agreement is reached.
 We find out the characteristic diffractive patterns at both RHIC and LHC energies and predict the
 impact parameter dependent $\cos 2\phi$ azimuthal asymmetries for $\rho^0$ photoproduction by considering UPCs and peripheral
 collisions. The future experimental measurements at RHIC and LHC relevant to our calculations will
 provide a  tool to rigorously investigate the coherent and incoherent production of vector meson in UPCs,
 as well as to probe the nuclear structure in heavy ion collisions.}

\begin{document}
\maketitle
\flushbottom

 \section{Introduction}

 Ultraperipheral collisions (UPCs) of heavy ions at the Relativistic Heavy Ion Collider (RHIC) and the
 Large Hadron Collider (LHC) offer a great opportunity to explore nuclear structure with beams of quasi-real
  photons before the Electron Ion Collider (EIC) era. In UPCs the strong hadronic interaction is suppressed,
   and the photon-nucleus ($\gamma$A) interactions involving photons emitted from one of the colliding nuclei are
    expected to be dominant. Due to the large flux of quasi-real photons, $\gamma$A interactions are enhanced
    by a factor $Z^2$ as compared to those in proton-nucleus (pA) or electron-nucleus (eA) collisions
     where $Z$ is the nuclear charge number. Among many exciting directions of UPC studies, see
     for example \cite{Hagiwara:2017fye,Hatta:2019ocp}, diffractive vector mesons photoproduction
     on nuclei provide access to the three dimensional gluon tomography of nucleus as well as stringent
     tests of the color glass condensate (CGC) description of saturation physics. Because of this,
     such processes have been extensively studied from both
     theoretical~\cite{Donnachie:1987pu,Ryskin:1992ui,Brodsky:1994kf,Nemchik:1996cw,Klein:1999qj,Munier:2001nr, Kopeliovich:2001xj,Kowalski:2003hm,Kowalski:2006hc,Rebyakova:2011vf,Guzey:2013qza,Lappi:2010dd,Guzey:2016piu,Xie:2016ino,Cai:2020exu}
     and experimental~\cite{Breitweg:1998nh,Chekanov:2002xi,Khachatryan:2016qhq,Adamczyk:2017vfu,STAR:2019yox,Sirunyan:2019nog,Acharya:2020sbc} sides during the past few decades.

Recently, significant $\cos 2\phi$ and $\cos 4\phi$ asymmetries for  $\rho^0$ meson production
 in UPCs have been observed by STAR collaboration~\cite{Daniel},
 where $\phi$ is the angle between the produced $\rho^0$ meson's transverse momentum and
 its decay product pion's transverse momentum. As the angular distribution of final state decayed pions
 contains the information of the polarization of $\rho^0$, the observed angular correlation between $\rho^0$
 and pion can be converted into the correlation between the transverse spin vector and the transverse
 momentum for $\rho^0$, thus the $\phi$ asymmetry can  serve as the meson's spin analyzer.
 The investigations of such polarization dependent observable in vector meson production certainly open
 a  new window to study the small-$x$ structure of heavy nuclei as well as the associated nontrivial QCD
 dynamics.

Motivated by the recent measurement by STAR collaboration at
RHIC~\cite{Daniel}, we carry out a detailed analysis of the $\cos
2\phi$ azimuthal  asymmetry for diffractive meson production in
UPCs. The underlying physics of $\cos 4\phi$ asymmetry is rather
different and will be addressed in a future work. Our calculation is
formulated in a conventional method: the quasi-real photon is
treated as the color dipole of a quark-antiquark pair which
recombines  to form a vector meson after scattering off the CGC
state inside a nucleus. To account for the $\cos 2\phi$ asymmetry
within the dipole model, the key insight is that the incident photon
is highly linearly polarized along the direction of its transverse
momentum. The correlation between the initial state photon's
polarization and transverse momentum will be transferred to that for
the final state vector meson. Notice that the Eikonal approximation
employed in the dipole approach plays a crucial role in preserving
spin information after the quark-antiquark
 pair experiences multiple gluon re-scattering.

As a matter of fact, the gauge bosons (photons/gluons) being highly
linearly polarized in the small $x$ limit have been recognized as a
common feature of the gauge theories in a series of
publications~\cite{Metz:2011wb, Li:2019yzy,Li:2019sin}. It was shown
in
Refs.~\cite{Metz:2011wb,Dominguez:2011br,Boer:2010zf,Akcakaya:2012si,Dumitru:2015gaa,Kotko:2015ura,Boer:2017xpy,Marquet:2017xwy,Gutierrez-Reyes:2019rug,Scarpa:2019fol,Kishore:2018ugo}
that the linear polarization of photons/gluons can be probed through
the azimuthal asymmetries in two particles correlations. For
instance, the QED calculations~\cite{Li:2019yzy,Li:2019sin} predict
a sizable $\cos 4\phi$ azimuthal asymmetry for pure electromagnetic
dilepton production in heavy ion collisions. Such $\cos 4\phi$
modulation has been clearly seen in a recent STAR
measurement~\cite{Adam:2019mby}. In particular, the computed impact
parameter dependent asymmetry is in excellent agreement with the
experimental data for the UPC case, while the QED calculation in
peripheral collisions slightly overestimates the asymmetry in the
centrality region $60-80\%$. With it being experimentally confirmed,
the linearly polarized quasi-real photon beam in heavy ion
collisions can be used as a powerful tool to explore the novel QCD
phenomenology as well. The current work represents the first effort
towards this direction.

The paper is organized as follows. In Sec.~\ref{sec-setup}, we
derive the joint impact parameter and transverse momentum dependent
cross section  in UPCs including both the coherent and incoherent
vector meson photoproduction contributions. In Sec.~\ref{sec-num},
we present numerical estimations of polarization averaged and $\cos
2\phi$ azimuthal asymmetries for diffractive $\rho^0$ production at
RHIC and LHC energies. Reasonable good agreement with the STAR
measurements are reached.  Finally,  the paper is summarized in
Sec.~\ref{sec-sum}.

 \section{Theoretical setup}
 \label{sec-setup}
 \subsection{The polarization dependent  wave functions}

 \begin{figure}[hbt]\centering
\includegraphics[angle=0,scale=0.6]{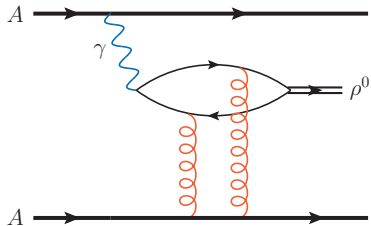}
\caption{ Diagram for diffractive $\rho^0$ meson production in ultraperipheral heavy ion collisions.}
\label{fig-diagram}
\end{figure}
In this paper, we consider vector meson $\rho^0$ production in UPCs,
$A+A \to \rho^0 + A' +A'$. In this process, as shown in Fig.
\ref{fig-diagram}, one of the nuclei can be considered as the source
of quasi-real photons that scatter off the other nucleus. The
quasi-real photon-nucleus interaction is treated as the
quark-antiquark color dipole scattering off the target nucleus in
the dipole picture. After the dipole-nucleus collision,
quark-antiquark pair subsequently recombines to form  a vector
meson. The calculation of the polarization averaged cross sections
for both the coherent and incoherent vector meson  production in
UPCs are well formulated within the dipole model in the
literatures~\cite{Ryskin:1992ui,Brodsky:1994kf}. Extending the
analysis to the polarization dependent case is the main purpose of
the present work.  At high energy, the transverse positions of the
quark and antiquark are not altered in the scattering process under
the eikonal approximation. Thus the production amplitude $ {\cal
A}(\Delta_\perp)$ can be conventionally expressed as the convolution
of the dipole scattering amplitude and the overlap between the
vector meson and photon wave functions in position space, \bea
 {\cal A}(\Delta_\perp)=i \!\int \! d^2
b_\perp e^{i \Delta_\perp \cdot b_\perp} \!\int \! \frac{d^2
r_\perp}{4\pi} \! \int_0^1 \!\! dz \  \Psi^{\gamma\rightarrow q\bar q}
 (r_\perp,z,\epsilon_{\perp}^{\gamma})N(r_\perp,b_\perp)
\Psi^{ V\rightarrow q \bar q *} (r_\perp,z,\epsilon_{\perp}^{V}),
\eea where $-\Delta_\perp$ is the nucleus recoil transverse
momentum. $\epsilon_{\perp}^{\gamma}$ and $\epsilon_{\perp}^{V}$ are
the magnitudes of transverse polarization vectors for the incident
quasi-real photon and final outgoing vector meson, respectively. The
polarization dependent wave function $\Psi^{\gamma\rightarrow q\bar
q}$ ($\Psi^{ V\rightarrow q \bar q}$ ) of the quasi-real photon
(vector meson) is determined from light cone perturbation theory at
leading order in the section below. $z$ denotes the fraction of the
photon's light-cone momentum carried by the quark.
$N(r_\perp,b_\perp)$ is the elementary amplitude for the scattering
of a $q\bar q$ dipole of size $r_\perp$ on a target nucleus at the
impact parameter $b_\perp$ of the $\gamma A$ collision.

For  coherent vector meson production, the dipole interacts with the
nucleus as a whole and leaves the nucleus in the ground state after
the collision. As a comparison, in the incoherent production process
the photon interacts with a nucleon inside the nucleus to produce a
vector meson leaving the nucleus in an excited state. The coherent
cross section is obtained by averaging the amplitude over the
position of the nucleon in the nucleus before squaring it $|\langle
{\cal A} \rangle_N|^2$, while the incoherent one is given by the
variance $\langle |{\cal A}|^2\rangle_N-|\langle {\cal
A}\rangle_N|^2$. Following Refs.
\cite{Kopeliovich:2001xj,Lappi:2010dd}, the incoherent production
amplitude squared (neglecting nuclear correlation) takes the form,
\bea | {\cal A}(\Delta_{\perp})|^2_{in} \approx & A(2 \pi B_p )^2
e^{-B_p\Delta_\perp^2}   \int d^2 b_\perp T_A(b_\perp) \bigg |  \int
\frac{d^2 r_\perp}{4\pi} \int_0^1 \!\! dz \Psi^{\gamma\rightarrow
q\bar q} (r_\perp,z,\epsilon_{\perp}^{\gamma})
\nonumber \\
&\times \Psi^{V\rightarrow q \bar q  *}
(r_\perp,z,\epsilon_{\perp}^{V}) {\cal N}(r_\perp)e^{-2\pi (A-1)B_p
T_A(b_\perp) {\cal N}(r_\perp) } \bigg |^2, \label{eq2} \eea where
$A$ is the nuclear atomic number and $B_p=4 ~\text{GeV}^{-2}$ in the
IPsat model~\cite{Kowalski:2003hm,Kowalski:2006hc}. $T_A(b_\perp)$
is the nuclear thickness function. ${\cal N}(r_\perp)$ is the
dipole-nucleon scattering amplitude. Eq. (\ref{eq2}) has a clear
physical interpretation: The dipole scatters independently off the
nucleons inside a nucleus, whose distribution in the transverse
plane is given by $T_A(b_\perp)$, and the dipole can further
interact with the rest of the $A-1$ target nucleons. While only
elastic interactions are allowed in diffractive process, the
inelastic re-scattering would make the process not diffractive and,
hence, should be rejected. The probability of not having inelastic
scattering is given by the factor $e^{-2\pi (A-1)B_p T_A(b_\perp)
{\cal N}(r_\perp) }$.

We now move on to work out the polarization dependent photon's wave
function.  For an ultrarelativistic charged heavy ion, the dominant
component of the induced electromagnetic gauge potential is the plus
component. The wave function of such a longitudinally polarized
photon can be perturbatively calculated directly. Alternatively, by
invoking the Ward identity argument, one can derive the same wave
function with polarization vector $-k_\perp^\mu/x$ instead of
$P^\mu$ for a quasi-real photon that carries momentum
$xP^\mu+k_\perp^\mu$~\cite{Li:2019sin}, where $P^\mu$ is the
four-momentum for the beam nucleus. This actually is an essential
reason why the small $x$ photons/gluons are highly linearly
polarized for a given $k_\perp$ in the  TMD description of
photon/gluon distributions. The forward polarization dependent  wave
function at leading order reads \bea \Psi^{\gamma \rightarrow q\bar
q} (r_\perp,z,\epsilon_\perp^\gamma) =&\frac{e e_q}{2\pi}
\delta_{aa'} \Bigg \{ \delta_{\sigma,- \sigma'} \left [  (1-2z)i
  \epsilon_{\perp}^\gamma \! \cdot r_\perp+\sigma \epsilon_{\perp}^\gamma \! \times r_\perp\right ]
 \frac{-1}{|r_\perp|} \frac{\partial}{\partial |r_\perp|}
\nonumber \\
& +\delta_{\sigma \sigma'} m_q(\epsilon_\perp^{\gamma,1}+i\sigma \epsilon_\perp^{\gamma,2}) \Bigg \}
 K_0(|r_\perp| e_f),
 \label{eq-wave-gamma}
\eea
where $\epsilon_{\perp}^\gamma=\hat k_\perp \equiv k_\perp/|k_\perp|$. And $\sigma$ and $\sigma'$ are the quark and antiquark helicities, $a$ and $a'$ are their color indices. $m_q$ and $e_q$ denote the quark mass and quark's electric charge number with flavor $q$, $e$ is the charge of the nucleus. $K_0$ is a modified Bessel function of the second kind, in its argument $e_f$ is defined as $e_f^2=Q^2z(1-z)+m_q^2$ with $Q^2 = k_\perp^2+x^2 M_p^2$, where $M_p$ is the proton mass.

In analogy to the  photon wave function, the forward transversely polarized vector meson wave function is given by~\cite{Kowalski:2003hm,Kowalski:2006hc},
\bea
\Psi^{V\rightarrow q \bar q} (r_\perp,z,\epsilon_\perp^{V})
=& \delta_{aa'}
\Bigg \{ \delta_{\sigma,- \sigma'} \left [  (2z-1)i
  \epsilon_{\perp}^V \cdot r_\perp+\sigma \epsilon_{\perp}^V \times r_\perp\right ]
 \frac{-1}{|r_\perp|} \frac{\partial}{\partial |r_\perp|}
 \nonumber \\
+& \delta_{\sigma \sigma'} m_q(\epsilon_\perp^{V,1}+i\sigma
\epsilon_\perp^{V,2}) \Bigg \} \Phi(|r_\perp|,z) \label{eq-wave-V}
\eea where the scalar part $\Phi(|r_\perp|,z)$ will be specified
shortly.

Combining Eqs.~(\ref{eq-wave-gamma}) and (\ref{eq-wave-V}), and
summing over the color and helicities of the quark and antiquark, we
obtain the overlaps between the photon and the vector meson wave
functions, \bea \sum_{a,a',\sigma,\sigma'} \Psi^{\gamma \rightarrow
q\bar q} \Psi^{V\rightarrow q \bar q  *} = & \frac{e e_q}{\pi} N_c
e^{i(z-\frac{1}{2} )\Delta_\perp \cdot r_\perp} \Bigg \{
\frac{1}{r_\perp^2} \left [ \frac{\partial}{\partial |r_\perp|}
\Phi^*(|r_\perp|,z) \right]
 \left [ \frac{\partial}{\partial |r_\perp|} K_0(|r_\perp| e_f) \right ]
 \nonumber \\
 \times& \left [ (2z-1)^2 (\epsilon_{\perp}^{V*} \! \cdot r_\perp)(\epsilon_{\perp}^\gamma \! \cdot r_\perp) +
 (\epsilon_{\perp}^{V*} \times r_\perp) (\epsilon_{\perp}^\gamma \times r_\perp) \right ]
 \nonumber \\
 + & m_q^2(\epsilon_{\perp}^\gamma \! \cdot \epsilon_{\perp}^{V*})
 \Phi^*(|r_\perp|,z)K_0(|r_\perp| e_f) \Bigg \}.
\eea Note that a phase factor  $ e^{i(z-\frac{1}{2} )\Delta_\perp
\cdot r_\perp}$ is included to  account for the non-forward
correction~\cite{Bartels:2003yj,Hatta:2017cte} (see also the
application of this phase factor in a model calculation in
proton-proton elastic scatterings~\cite{Hagiwara:2020mqb}). As we
focus on low transverse momentum region where the produced meson
transverse momentum is of the order of  $1/R_A$ with $R_A$ the
nuclear radius, $\Delta_\perp $ is sufficiently small compared to
the relevant value of $1/r_\perp$. Therefore we will neglect the
phase $ e^{i(z-\frac{1}{2} )\Delta_\perp \cdot r_\perp}$ to further
simplify the expression. By doing so, the overlap of photon and
meson wave functions  can be cast into the following form after
integrating out the azimuthal angle of $r_\perp$, \bea
\sum_{a,a',\sigma,\sigma'} \Psi^{\gamma \rightarrow q\bar q}
\Psi^{V\rightarrow q \bar q  *} = &  (\epsilon_{\perp}^{V*}  \cdot
\epsilon_{\perp}^\gamma ) \frac{e e_q}{2\pi} 2N_c  \int \frac{d^2
r_\perp}{4\pi}  \ N(r_\perp,b_\perp) \Bigg \{ \left [ z^2+(1-z)^2
\right ]
\nonumber \\
 \times &
\frac{\partial\Phi^*(|r_\perp|,z)}{\partial |r_\perp|}
 \frac{\partial  K_0(|r_\perp| e_f)}{\partial |r_\perp|}
  +m_q^2 \Phi^*(|r_\perp|,z)K_0(|r_\perp| e_f) \Bigg \},
 \label{eq-wave-combine}
\eea where the correlation between $r_\perp$ and $b_\perp$ in
$N(r_\perp,b_\perp)$ is ignored~\cite{Hatta:2016dxp,Zhou:2016rnt}.
In Eq.~(\ref{eq-wave-combine}), it can be clearly seen that the
photon's polarization vector manifestly couples to meson's one. As
mentioned in the introduction, the coupling of the spin states is
the consequence of the Eikonal approximation employed in our
calculation.

\subsection{The polarization dependent differential cross section}
The purpose of the current work is to investigate the angular
correlation  between the vector meson's transverse spin vector and
its decayed pion's transverse momentum. At leading order in
perturbative QCD, the meson's transverse momentum is equal to the
sum of the incident photons' transverse momentum $k_\perp$ and
$\Delta_\perp$. It is then natural to formulate the transverse
momentum dependent cross section in the framework of TMD
factorization, which reads
\begin{eqnarray}
\frac{d \sigma}{d^2 q_\perp dY}=\frac{1}{4\pi^2} \int d^2
\Delta_\perp d^2k_\perp x f(x,k_\perp)
\delta^2(k_\perp+\Delta_\perp-q_\perp) \langle | {\cal
A}|^2\rangle_N,
\end{eqnarray}
where $q_\perp$ and $Y$ are the produced vector meson's transverse momentum and rapidity, respectively. The photon TMD distribution is denoted as $f(x,k_\perp)$ which will be computed below using the equivalent photon approximation, where longitudinal momentum fraction $x$ is fixed as $x= \sqrt{\frac{q_\perp^2+M_V^2}{s}}e^{Y}$ at leading order. Correspondingly, the longitudinal momentum fraction transferred to the vector meson via the dipole-nucleus interaction is given by  $x_g= \sqrt{\frac{q_\perp^2+M_V^2}{s}}e^{-Y}$.

We proceed by explicitly separating the coherent and incoherent contributions,
\bea
\frac{d \sigma}{d^2 q_\perp dY} =& \frac{{\cal C}}{4\pi^2} \int d^2 \Delta_\perp d^2k_\perp x f(x,k_\perp)
\delta^2(k_\perp + \Delta_\perp - q_\perp)(\epsilon_{\perp}^{V*} \cdot \hat k_\perp )^2
\nonumber \\
 \times &
 \left [ |{\cal A}_{co}(\Delta_\perp)|^2\!+\int d^2 b_\perp T_A(b_\perp)
 |{\cal A}_{in}(\Delta_\perp)|^2 \right ]
 \nonumber \\
 =&\frac{{\cal C}}{8\pi^2}
 \int d^2 \Delta_\perp  x f(x,q_\perp-\Delta_\perp)\bigg\{1+\cos 2\phi \left [ 2(\hat q_\perp \cdot \hat k_\perp )^2 - 1 \right ]\bigg\}
 \nonumber \\
\times & \left [ |{\cal A}_{co}(\Delta_\perp)|^2+\int d^2 b_\perp T_A(b_\perp)|{\cal A}_{in}(\Delta_\perp)|^2 \right ],
 \label{intcs}
\eea
where $\hat q_\perp=q_\perp/|q_\perp|$ and  $\phi$ is the angle between $\epsilon_{\perp}^{V*}$ and $q_\perp$. We have replaced $\epsilon_\perp^\gamma$ with $\hat k_\perp$ in the above formula.  A pre-coefficient ${\cal C}$ is introduced here to account for the real part of the amplitude as well as the skewedness effect. In our numerical estimations, we simply neglect these effects and set ${\cal C}$ to be equal to 1. In Eq. (\ref{intcs}), the coherent and incoherent scattering amplitudes are respectively given by
\bea
 {\cal A}_{co}(\Delta_\perp) =& \int d^2
b_\perp e^{-i \Delta_\perp \cdot b_\perp} \int \frac{d^2
r_\perp}{4\pi}  \ N(r_\perp,b_\perp) [\Phi^*K](r_\perp),
 \\
 {\cal A}_{in}(\Delta_\perp) =& \sqrt{A }2 \pi B_p e^{-B_p\Delta_{\perp}^2/2}
    \left [  \int \frac{d^2
r_\perp}{4\pi}  {\cal N}(r_\perp)e^{-2\pi (A-1)B_p T_A(b_\perp)
{\cal N}(r_\perp) } [\Phi^*K](r_\perp) \right ], \eea where
$[\Phi^*K]$ denotes the overlap of the virtual photon wave function
and the vector meson wave function, \bea [\Phi^*K](r_{\perp}) =&
\frac{N_{c} e e_q}{\pi} \int_0^1 dz \bigg \{m_q^2
\Phi^*(|r_\perp|,z)K_0(|r_\perp| e_f) + \left [ z^2 + (1 - z)^2
\right ]
\nonumber \\
\times &
 \frac{\partial\Phi^*(|r_\perp|,z)}{\partial |r_\perp|}
 \frac{\partial  K_0(|r_\perp| e_f)}{\partial |r_\perp|} \bigg \}.
\eea

It is now worthwhile to point out that the impact parameter $\tilde
b_\perp$ of the two colliding nuclei is implicitly integrated out in
Eq.~(\ref{intcs}). Therefore, one can not compute the observables in
UPCs using Eq.~(\ref{intcs}). It is necessary to introduce an impact
parameter $\tilde b_\perp$ dependent cross section, from which  the
UPC observables can be estimated by integrating out $\tilde b_\perp$
from $2 R_A$ to $\infty$. Such a formalism actually has been
developed long ago in the context of evaluating the electromagnetic
dilepton production in UPCs~\cite{Vidovic:1992ik,Hencken:1993cf}.
Previously, the $\tilde b_\perp$ dependent azimuthal asymmetries for
dilepton production was studied following the same
method~\cite{Li:2019sin}.

\begin{figure}[hbt]\centering
\includegraphics[angle=0,scale=0.6]{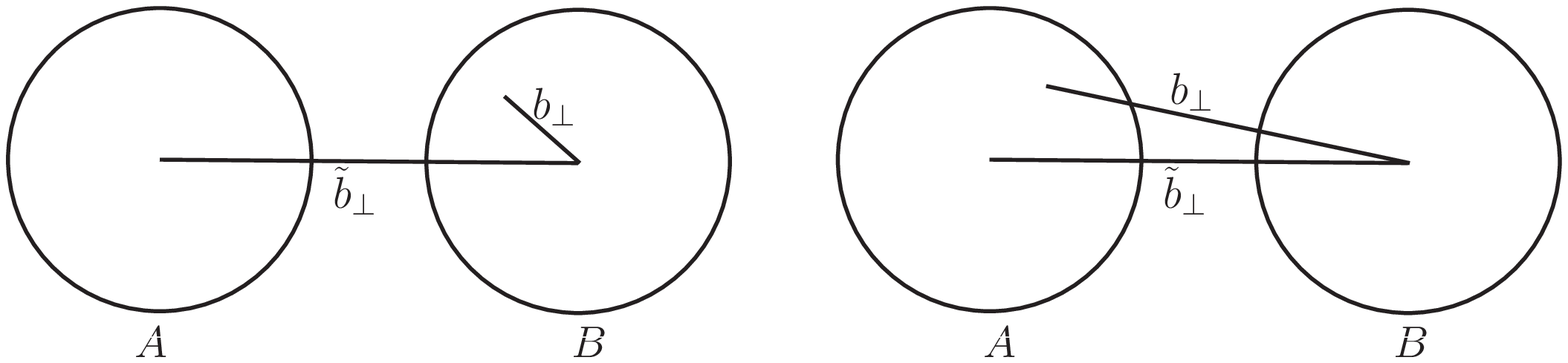}
\caption{ The vector meson is  locally produced  in the transverse
plane  inside each nucleus(A or B) which takes turns to act as the
target and the quasi-real photon source. This creates a set up of
the Young's double-slit experiment at fermi scale. To suppress
hadronic interactions, $\tilde b_\perp$ must be larger than $2
R_A$.} \label{fig-bdep}
\end{figure}

The precise determination of the joint transverse momentum and
impact parameter dependence crucially relies on the assumption that
the lepton pair or vector meson is locally produced in the
transverse plane of nucleus. This requirement is satisfied as long
as the vector meson's mass is much larger than the inverse of the
nucleus radius. The probability amplitude for coherently  producing
a meson at the position $b_\perp$ inside two nuclei is then
proportional to, \bea
   {\cal M}(Y,\tilde b_\perp,b_\perp) \propto
 \left [ {F}_A(Y, b_\perp- \tilde b_\perp )
N_B(Y,b_\perp)
 + N_A(-Y,b_\perp - \tilde b_\perp){F}_B(-Y, b_\perp) \right ],
 \eea
where  $F_B$ is the EM gauge potential induced by nucleus $B$. The
$r_\perp$ dependence of the dipole amplitude $ N_A(Y,b_\perp)$ is
suppressed for brevity. Note that  each incident ion can serve as a
source of photons and a target. So the production amplitude contains
two contributions, shown in Fig.~\ref{fig-bdep}, corresponding to
the right-moving photon source (denoted as nucleus $A$) and the
left-moving source (denoted as  nucleus $B$).  Since these two
possibilities are indistinguishable, they should be summed up on the
amplitude level rather than on the cross section level.

We now Fourier transform the above expression to momentum space,
\bea
   {\cal M}(Y,\tilde b_\perp, q_\perp)&\! =\!\int d^2 b_\perp e^{-ib_\perp \cdot q_\perp}
{\cal M}(Y,\tilde b_\perp, b_\perp)
   \propto \! \int \frac{d^2 k_\perp}{(2\pi)^2}
   \frac{d^2 \Delta_\perp}{(2\pi)^2}(2\pi)^2 \delta^2(q_\perp-\Delta_\perp-k_\perp)
\nonumber \\&\times
   \bigg \{ {F}_A(Y, k_\perp) N_B(Y,\Delta_\perp)
    e^{-i\tilde b_\perp \cdot k_\perp}
  + {F}_B(-Y, k_\perp) N_A(-Y,\Delta_\perp) e^{-i\tilde b_\perp \cdot \Delta_\perp}  \bigg \},
\eea
where a nontrivial phase arises  together with a normal delta function which ensures transverse momentum conservation. Due to the different phase factors $e^{-i\tilde b_\perp \cdot k_\perp} $ and $e^{-i\tilde b_\perp \cdot \Delta_\perp}$, a large destructive interference could  occur between two contributions  as  shown below. Such destructive interference of $\rho^0$ in UPCs was first proposed by Klein and Nystrand~\cite{Klein:1999gv}, and verified by the STAR measurement~\cite{Abelev:2008ew}.  Later, the authors of the paper~\cite{Zha:2018jin} suggested that this phenomenon could also be studied in hadronic heavy ion collisions.

After combining with the conjugate amplitude, it  yields  phases
$e^{\pm i\tilde b_\perp \cdot (k_\perp-k_\perp')}$ for the diagonal
terms and
  $e^{\pm i\tilde b_\perp \cdot (\Delta_\perp-k_\perp')}$  for the interference term,
  where $k_\perp'$ is the photon's transverse momentum in the
 conjugate amplitude, which is not necessarily identical to that in the amplitude.
 One eventually ends up with the joint $\tilde b_\perp$ and $q_\perp$ dependent cross section,
\bea
\frac{d \sigma}{d^2 q_\perp dY d^2 \tilde b_{\perp} }
=& \frac{1}{(2\pi)^4} \int d^2 \Delta_\perp d^2k_\perp d^2 k_\perp'
\delta^2(k_\perp+\Delta_\perp-q_\perp)
 (\epsilon_{\perp}^{V*}  \cdot \hat k_\perp )(\epsilon_{\perp}^{V} \cdot \hat k_\perp' )
 \bigg \{
\int d^2   b_\perp
 \nonumber \\
\times&  e^{i \tilde b_\perp \cdot (k_\perp'-k_\perp)}
\left [ T_A(b_\perp) {\cal A}_{in}(Y,\Delta_\perp) {\cal A}_{in}^*(Y,\Delta_\perp')
{\cal F}(Y,k_\perp){\cal F}(Y,k_\perp')
  + ( A \leftrightarrow B)
\right ]
 \nonumber \\
+ &\left [  e^{i \tilde b_\perp \cdot (k_\perp' -k_\perp)}
{\cal A}_{co}(Y,\Delta_\perp) {\cal A}_{co}^*(Y, \Delta_\perp')
{\cal F}(Y,k_\perp){\cal F}(Y,k_\perp')
 \right ]
  \nonumber \\
+ & \left [  e^{i \tilde b_\perp \cdot (\Delta_\perp'-\Delta_\perp)}
{\cal A}_{co}(-Y,\Delta_\perp) {\cal A}_{co}^*(-Y, \Delta_\perp')
{\cal F}(-Y,k_\perp){\cal F}(-Y,k_\perp')
 \right ]
\nonumber \\
+&  \left [ e^{i \tilde b_\perp \cdot (\Delta_\perp' -k_\perp)}
 {\cal A}_{co}(Y,\Delta_\perp) {\cal A}_{co}^*(-Y, \Delta_\perp'){\cal F}(Y,k_\perp){\cal F}(-Y,k_\perp')
 \right ]
    \nonumber \\
  + &  \left [ e^{i \tilde b_\perp \cdot (k_\perp' -\Delta_\perp)}
 {\cal A}_{co}(-Y,\Delta_\perp) {\cal A}_{co}^*(Y, \Delta_\perp'){\cal F}(-Y,k_\perp){\cal F}(Y,k_\perp')
 \right ]
  \bigg \},
   \label{fcs}
 \eea
where ${\cal F}(Y,k_\perp)$ is related to the coherent photon TMD
via the relation $\left [{\cal F}(Y,k_\perp) \right ]^2=x
f(x,k_\perp)$, and will be specified shortly. $\Delta_\perp'$  is
constrained by the transverse momentum conservation:
$k_\perp+\Delta_\perp=k_\perp'+\Delta_\perp'$. The  diagonal term
and the interference term from the coherent production contribution
are presented in the last four lines. The incoherent production
contribution is given in the second line, where the interference
term is ignored due to its smallness at low transverse momentum. To
demonstrate the destructive interference effect, one can carry out
$\tilde b_\perp$ integration and obtains the delta function
$\delta^2(k_\perp-k_\perp')$ associated with the diagonal term and
$\delta^2(\Delta_\perp-k_\perp')$ for the interference term
\footnote{It can be readily seen that the $\tilde b_\perp$
integrated cross section is reduced to Eq.~(\ref{intcs}) provided
that the interference term is neglected.}. It now becomes evident
that two contributions at $q_\perp=0$ have an opposite sign
resulting from the vector product structure $ (\epsilon_{\perp}^{V*}
\! \cdot \hat k_\perp )(\epsilon_{\perp}^{V} \! \cdot \hat k_\perp'
)$. For the fully symmetrical case $Y=0$, this effect leads to a
complete cancelation between the last four lines at $q_\perp=0$.
Such cancelation can be intuitively  understood as the consequence
of the parity conservation. In the general case without $\tilde
b_\perp$ integration, the cross section is reduced by this
destructive interference effect mainly in the low $q_\perp$ region.

To facilitate the numerical estimation, we replace the vector product structure
 $ (\epsilon_{\perp}^{V*} \! \cdot \hat k_\perp )(\epsilon_{\perp}^{V} \! \cdot \hat k_\perp' )$
 in Eq.~(\ref{fcs}) with~\cite{Boer:2009nc},
\begin{eqnarray}
\left [
(\hat k_\perp \! \cdot \hat k_\perp' )+ \cos(2\phi)
\left (2(\hat k_\perp \cdot \hat q_{\perp})(\hat k_\perp' \cdot \hat q_{\perp})-\hat k_\perp \! \cdot \hat k_\perp' \right )
  \right ],
  \label{eq-decomps}
\end{eqnarray}
where the polarization states of the produced $\rho^0$ have been summed over. We now argue that the $\cos 2\phi$ asymmetry under investigation is essentially equivalent to the measured angular correlation between $q_\perp$ and  the final state pion's transverse momentum $p_\perp^{\pi}$. Due to the angular momentum conservation, the decay amplitude of the process $\rho^0 \rightarrow \pi^+ \pi^-$ must be proportional to ${\cal M} \propto e^{i \lambda \phi_{\pi}}$ where $\phi_{\pi}$ is the azimuthal angle of $p_\perp^{\pi}$ and $\lambda$ denotes $\rho$ meson's helicity state. This immediately implies that there exits  a angular correlation of the type $\hat p_\perp^{\pi} \cdot \epsilon_\perp^{V*}$ provided that the vector meson is linearly polarized.  As a consequence, once summing over all polarization states of the vector meson, the correlation  $ 2(\hat q_\perp \cdot \epsilon_\perp^{V*})^2-1 $ appears  in the above cross section formula will be converted into the one $ 2(\hat q_\perp \cdot \hat p_\perp^{\pi})^2-1 $ which is exactly the observable that has been measured by the STAR experiment.

\section{Phenomenology}
\label{sec-num} We proceed to perform the numerical estimations of
the $\cos 2\phi$ asymmetry using Eq.~(\ref{fcs}) in this section.
First of all, let us collect all ingredients that are necessary for
numerical calculations. We start with introducing the
parametrization for the dipole scattering amplitude whose formal
operator definition is given by,
\begin{eqnarray}
N(b_\perp, r_\perp)=1-  \frac{1}{N_c} \left \langle
{\rm Tr} \left( U(b_\perp+r_\perp/2) U^\dag(b_\perp-r_\perp/2) \right ) \right \rangle.
\end{eqnarray}
The dipole amplitude is usually obtained by solving the BK equation
with the initial condition being fitted to the experimental data or
derived from the MV model. However, the numerical implementation of
the impact parameter dependent BK equation is a highly non-trivial
task. For simplicity,  we instead use a phenomenological
parametrization for the $b_\perp$ dependence of the dipole
amplitude~\cite{Kowalski:2003hm,Kowalski:2006hc},
\begin{eqnarray}
N(b_\perp, r_\perp)
=1-e^{-2\pi B_p A T_A(b_\perp) {\cal N}(r_\perp)},
\end{eqnarray}
where, as mentioned before, ${\cal N}(r_\perp)$ is the
dipole-nucleon scattering amplitude. The nuclear thickness function
$T_A(b_\perp)$ is determined with the Woods-Saxon distribution in
our numerical calculation. Note that $4\pi B_p{\cal
N}(r_\perp)=\sigma_{\text {dip}}^p(r_\perp)$ is the total
dipole-proton cross section for a dipole of size $r_\perp$. In
literatures there are many parameterizations available for the
dipole-proton cross section. Here we adopt  a modified IPsat model
in which the impact parameter dependence of the dipole-nucleon
scattering amplitude has been factorized out~\cite{Lappi:2010dd}
\begin{eqnarray}
{\cal N}(r_\perp)=1-\exp \left[-r_\perp^2 G(x_g,r_\perp) \right ] ,
\end{eqnarray}
where $G$ is proportional to the DGLAP evolved gluon distribution in
the Bartels, Golec-Biernat and Kowalski (BGBK)
parametrization~\cite{Bartels:2002cj}.  In our numerical
estimations, we adopt a simpler parametrization for the gluon
distribution known as the Golec-Biernat and W\"usthoff (GBW)
model~\cite{Golec-Biernat:1998js,Golec-Biernat:1999qd},
\begin{eqnarray}
G(x_g)= \frac{1}{4}Q_s^2(x_g),
\end{eqnarray}
where $Q_s(x_g)=(x_0/x_g)^{\lambda_{GBW}/2}$ GeV is the saturation
scale. We use the parameters $x_0=3\times10^{-4}$ and
$\lambda_{GBW}=0.29$~\cite{Kowalski:2006hc} which were determined by
fitting to HERA  data.

For the scalar part of  the vector meson wave function, we use
``Gaus-LC'' wave function also taken from
Refs.~\cite{Kowalski:2003hm,Kowalski:2006hc}
\begin{eqnarray}
\Phi(|r_\perp|,z)= \beta z(1-z) \exp
\left[-\frac{r_\perp^2}{2R_\perp^2}\right ],
\end{eqnarray}
where $\beta=4.47$, $R_\perp^2=21.9 ~\text{GeV}^{-2}$ for $\rho$
meson. An alternative parametrization, the ``boosted Gaussian'' wave
function is also widely used in the study of exclusive production of
vector meson. The existing HERA data is reasonably well described by
estimations of vector meson photoproduction employing either wave
function model.

The photon distribution $f(x,k_\perp)$ at low transverse momentum is
commonly computed with  the  equivalent photon approximation, also
often referred to as  the Weizs$\ddot{a}$cker-Williams method, in
which the photon flux is calculated by treating the fields of
charged relativistic heavy ions as external, i.e., classical
electromagnetic field. This method has been widely used to compute
UPC observables, see for example
Refs.~\cite{Klein:2018fmp,Zha:2018tlq}. The photon distribution
derived in the equivalent photon approximation is given
by~\cite{Bertulani:1987tz,Vidovic:1992ik}
\begin{eqnarray}
xf(x,k_\perp)=\frac{Z^2 \alpha_e}{\pi^2} k_\perp^2 \left [
\frac{F(k_\perp^2+x^2M_p^2)}{(k_\perp^2+x^2M_p^2)}\right ]^2,
\label{f1h1}
\end{eqnarray}
where  $F$ is the nuclear charge form factor, $M_p$ is the proton
mass. Similarly, one has ${\cal F}(Y,k_\perp)=\frac{Z \sqrt{
\alpha_e}}{\pi} |k_\perp|
\frac{F(k_\perp^2+x^2M_p^2)}{(k_\perp^2+x^2M_p^2)} $.
 The nuclear charge form factor is commonly determined with the
Woods-Saxon distribution,
\begin{eqnarray}
F(\vec k^2)= \int d^3 r e^{i\vec k\cdot \vec r} \frac{C^0}{1+\exp{\left [(r-R_{WS})/d\right ]}},
\end{eqnarray}
where  $R_{WS}$ (Au: 6.38 fm, pb: 6.62 fm) is the nuclear radius and $d$ (Au: 0.535 fm, Pb: 0.546 fm) is the skin depth, $C^0$ is the normalization factor. Alternatively, one can use the  form factor in momentum space from the STARlight MC generator~\cite{Klein:2016yzr},
\begin{eqnarray}
F(\vec k^2)=\frac{3}{|\vec k|^3 R_A^3}\left [ \sin(|\vec
k|R_A)-|\vec k|R_A \cos(|\vec k|R_A)\right ]\frac{1}{a^2 \vec k^2+1},
\label{ff}
\end{eqnarray}
where $R_A=1.1 A^{1/3}$ fm, and $a=0.7$ fm. This parametrization numerically is very close to the Woods-Saxon distribution, and will be used in our numerical evaluation. Due to the neutron skin effect and  the surrounding pion cloud, the effective nuclear strong interaction  radius is larger than its EM radius. To fit RHIC data~\cite{Adamczyk:2017vfu},  we compute the  thickness function $T_A(b_\perp)$ with the radius $R_A=6.9$ fm and the depth $d=0.64$ fm for a gold target. For a lead target, we simply re-scale these numbers by multiplying a factor $A_{\text{lead}}^{1/3}/A_{\text{gold}}^{1/3}$.  We  determine $e_q$ by noticing that the $\rho^0$ meson wave function reads $\frac{1}{\sqrt{2}}\left (|u \bar u \rangle-|d \bar d \rangle \right )$. This would imply a replacement of $e_q$  by $e_q \rightarrow \frac{1}{\sqrt{2}}(e_u-e_d)$.
The effective charge $e_q$ for $\rho^0$ then is $1/ \sqrt{2}$.

For the unrestricted UPC case,  the asymmetry is averaged over the
impact parameter range $[2R_A, \infty]$. However, RHIC-STAR measures
$\rho^0$ photoproduction cross section together with the double
electromagnetic excitation in both ions. Neutrons emitted at forward
angles from the scattered nuclei are detected by zero-degree
calorimeters (ZDCs), and used as a UPC trigger. Requiring that UPCs
are accompanied by forward neutron emission alters the impact
parameter distribution compared with unrestricted UPC events. In
order to incorporate the experimental conditions in the theoretical
calculations, one can define a ``tagged'' UPC cross section
\begin{eqnarray}
2 \pi \int_{2R_A}^{\infty} \tilde b_\perp d\tilde b_\perp P^2(\tilde b_\perp) d \sigma(\tilde b_\perp, \ ...).
\end{eqnarray}
Where the probability $P(\tilde b_\perp)$ of emitting a neutron from  the scattered nucleus is often parameterized as~\cite{Baur:1998ay}
\begin{eqnarray}
P(\tilde b_\perp)= P_{1n}(\tilde b_\perp)
\exp \left [-P_{1n}(\tilde b_\perp)\right ],
\end{eqnarray}
which is denoted as the ``1n'' event, while for emitting any number of neutrons (``Xn'' event), the probability is given by
\begin{eqnarray}
P(\tilde b_\perp)= 1-
\exp \left [-P_{1n}(\tilde b_\perp) \right ],
\end{eqnarray}
with $P_{1n}(\tilde b_\perp)= 5.45\times10^{-5} \frac{Z^3(A-Z) }{A^{2/3} \tilde b_\perp^2 } \ \text{fm}^2 $. As a matter of fact, the mean impact parameter is dramatically reduced in interactions with  Coulomb dissociation.

\begin{figure}[hbt]\centering
\includegraphics[angle=0,scale=0.9]{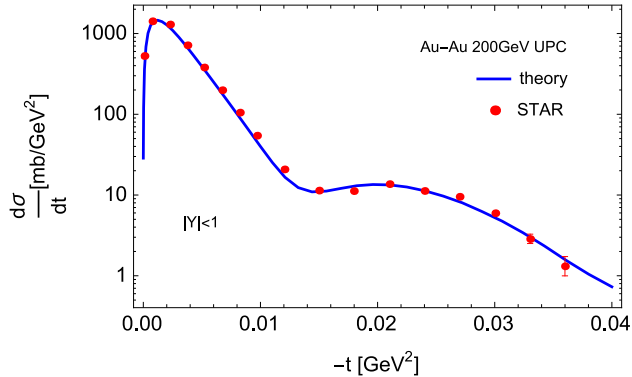}
\caption{ (color online) The unpolarized cross section for coherent
$\rho^0$ photo-production in XnXn events at RHIC energy.  The red
dots are experimental data points taken from \cite{Adamczyk:2017vfu}. The blue line shows
our numerical result for this unpolarized cross section. }
\label{unp}
\end{figure}

With all these ingredients, we are ready to  perform numerical study
of the azimuthal asymmetries for $\rho^0$ meson production in heavy
ion collisions.  To test the theoretical calculation, we first
compute the azimuthal averaged cross section for coherent
photoproduction of $\rho^0$ and compare with experimental data from
the STAR collaboration~\cite{Adamczyk:2017vfu}. In particular, we
calculate the differential cross section $d\sigma/dt$ with the
Mandelstam variable $t \approx -q_{\perp}^2$, and the rapidity is
integrated out in the region $|Y|\le 1$ to match the STAR
measurement. Notice that the incoherent component has been
subtracted out in STAR measurement. Therefore, we exclude the first
term in Eq.~(\ref{fcs}) and integrate over the azimuthal angle
$\phi$, namely only the first term in Eq.~(\ref{eq-decomps}) needs
to be considered. As shown in Fig.~\ref{unp}, our theoretical result
represented by blue curve describes the experimental data perfectly
in identifying the minima and peaks, as well as the overall shapes.

\begin{figure}[hbt]\centering
\includegraphics[angle=0,scale=0.75]{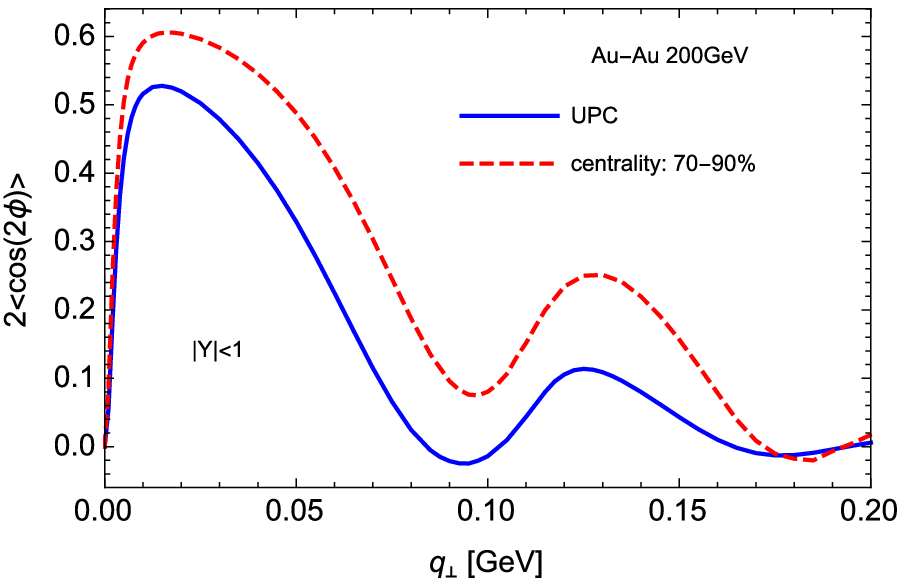}
\includegraphics[angle=0,scale=0.8]{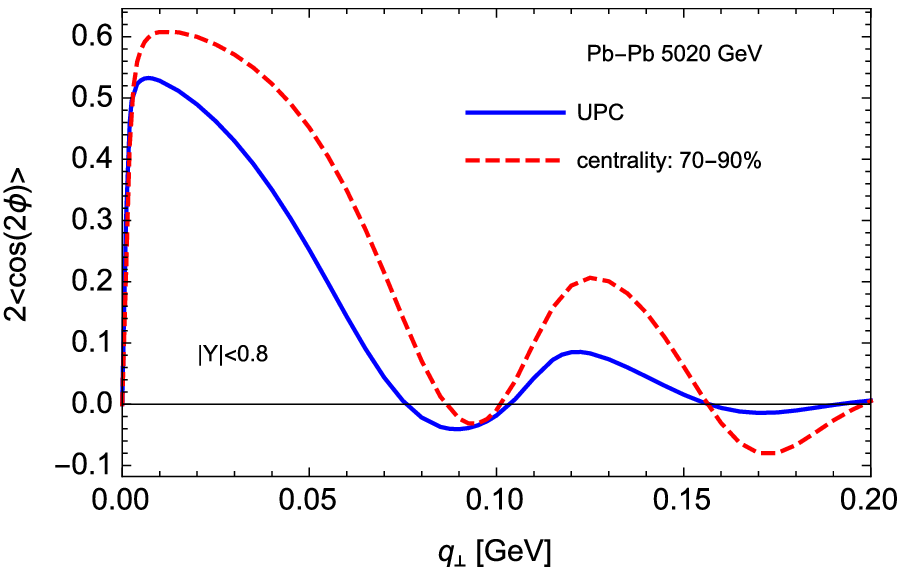}
\caption{(color online) The $\cos 2\phi$ azimuthal asymmetries in
$\rho^0$ production(Xn-Xn events) in heavy ion collisions at RHIC
and LHC energies. The computed $\cos 2\phi$ in UPC at RHIC
energy(left panel, solid line) can qualitatively describe  the
preliminary measurement by the STAR collaboration~\cite{Daniel}. The
asymmetry in peripheral collisions with centrality region from
70\%-90\% is also presented with the dashed lines.} \label{v2}
\end{figure}

The numerical results for the azimuthal asymmetries for  $\rho^0$ at
RHIC and LHC energies are presented in Fig.~\ref{v2}, where the
azimuthal asymmetry, i.e., the average value of  $\cos 2\phi$ is
defined as,
\begin{eqnarray}
\langle \cos(2\phi) \rangle &=&\frac{ \int \frac{d \sigma}{d {\cal PS}} \cos 2\phi \ d {\cal PS} }
{\int \frac{d \sigma}{d {\cal PS}}  d {\cal PS}}.
\end{eqnarray}
 We use exactly the same setups as that in the unpolarized case but
including both the coherent
  and incoherent components. Since we are considering the average value of  $\cos 2\phi$,
  only the second term in Eq.~(\ref{eq-decomps}) contributes.
   We can see clearly the diffractive pattern with two minima visible in $q_\perp$ distribution,
   such characteristic feature is also identified in the STAR preliminary measurement.
      The $q_\perp$ distribution for the average value of  $\cos 2\phi$ can be easily understood as
   the asymmetry is almost entirely generated in the coherent scattering,
   while both the coherent and incoherent production contribute to the azimuthal
   averaged cross section. At the relative large transverse momentum($q_\perp > 100 {\text MeV}$),
  most of  $\rho^0$ meson's transverse momentum originates from the nucleus.  Based on  this observation,
  one, after few steps of algebraic manipulations,  can show that the asymmetry is proportional to the slope of $\Delta_\perp$ distribution which gets very
  large near the first minima of the diffractive pattern. This can be clearly seen from our numerical
  result, where the second peak of the asymmetry located at the first minima of the unpolarized cross section.
   Nevertheless, as the first attempt,
   our result shown in the left plot in Fig.~\ref{v2}(solid line) describes the STAR preliminary data~\cite{Daniel} reasonably well
   in terms of finding the correct depths of the dips. However, slightly larger $q_{\perp}$ for
   the locations of the dips are found from our theoretical calculation comparing to those in
   STAR preliminary data, which suggests an increase in effective nuclear size in our calculation
   when considering polarized case. In order to investigate the impact parameter dependence,
   we also show  in Fig.~\ref{v2} the comparison between UPC and peripheral collisions
   at RHIC energy $\sqrt{s} = 200$ GeV in Au-Au collisions and at LHC energy $\sqrt{s}=5020$ GeV in
   Pb-Pb collisions, we take $70-90\%$ centralities as an example in peripheral collisions
   \footnote{For peripheral collisions with relative large impact parameter, the coherent photon-nucleus
   interaction still dominates over hadronic reactions in vector meson production~\cite{STAR:2019yox}.
   }.
   With the increase of impact parameter, we see slightly shift of the location for the dips.
   We also predict measurable difference between UPC and peripheral collisions at both RHIC and LHC.

\section{Conclusion}
\label{sec-sum} In summary, we have studied the $\cos 2\phi$
azimuthal angular correlation in vector meson production in
ultraperipheral heavy ion collisions, where $\phi$ is defined as the
angle between vector meson's transverse spin vector and its
transverse momentum. The asymmetry essentially results from the
linear polarization of incident coherent photons, which just has
been experimentally confirmed by the recent STAR measurement of a
$\cos 4\phi$ modulation in pure electromagnetic lepton pair
production~\cite{Adam:2019mby}. The asymmetries evaluated in the
dipole model for $\rho^0$ photoproduction at RHIC and LHC energies
are shown to be rather sizable. Admittedly, the perturbative
treatment for $\rho^0$  must be legitimately criticized due to the
lack of a hard scale in the problem. However, one might expect that
the angular correlation structure is not altered by the
non-perturbative effect, for which  a more sophisticated
phenomenological  method is required. Nevertheless, we found that
our calculation turns out to be in  reasonably good agreement with
the  $\rho^0$ measurement by STAR collaboration. As mentioned in the
introduction, a significant $\cos 4\phi$ asymmetry in $\rho^0$
production  was also observed at RHIC. This observable could
potentially give the access to the non-trivial gluon GTMD/Wigner
distribution and will be addressed in a future publication.

The obtained transverse momentum dependent $\cos 2\phi$ asymmetries
have a distinctive diffractive pattern which undoubtly opens a new
window to investigate the coherent and incoherent production of
vector meson.  As demonstrated by the present study, quasi-real
photon beams with linear polarization in heavy ion collisions can be
used as a powerful tool to explore novel QCD phenomenology.
Meanwhile, as a byproduct of this work, we developed a formalism to
compute the joint impact parameter and transverse momentum dependent
cross sections that enables us to reliably extract $\Delta_\perp$
dependence in UPCs. The Fourier transform of $\Delta_\perp$
distribution would provide crucial information on the transverse
spatial distribution of gluons inside a nucleus, which is one of the
central scientific goals in the forthcoming EIC era.

\begin{acknowledgments}
J. Zhou thanks Zhang-bu Xu, James Daniel Brandenburg, and Chi Yang
for helpful discussions. J. Zhou has been supported by the National
Science Foundations of China under Grant No.\ 11675093. Y. Zhou
has been supported by the National Science Foundations of China
under Grant No.\ 11675092. The work of H. X. is supported by the research startup funding at South China Normal University and Science and Technology Program of Guangzhou (No. 2019050001). CZ is
supported by the China Postdoctoral Science Foundation under Grant No.~2019M662317.
\end{acknowledgments}

{\bf Note added:} After the manuscript was posted on arXiv, a related preprint~\cite{Zha:2020cst} appeared soon, where
 the similar result for the azimuthal asymmetry was obtained.

\end {document}